# Predicting a Two-dimensional P$_2$S$_3$ Monolayer: A Global Minimum Structure

Hang Xiao[a], Xiaoyang Shi[a], Feng Hao[a], Xiangbiao Liao[a], Yayun Zhang[a,c] and Xi Chen*[a,b]

[a] Columbia Nanomechanics Research Center, Department of Earth and Environmental Engineering, Columbia Uni-versity, New York, NY 10027, USA, E-mail: xichen@columbia.edu

[b] SV Laboratory, School of Aerospace, Xi'an Jiaotong University, Xi'an 710049, China

[c] College of Power Engineering, Chongqing University, Chongqing 400030, China

**Abstract**: Based on extensive evolutionary algorithm driven structural search, we propose a new diphosphorus trisulfide (P$_2$S$_3$) 2D crystal, which is dynamically, thermally and chemically stable as confirmed by the computed phonon spectrum and *ab initio* molecular dynamics simulations. This 2D crystalline phase of P$_2$S$_3$ corresponds to the global minimum in the Born-Oppenheimer surface of the phosphorus sulfide monolayers with 2:3 stoichiometries. It is a wide band gap (4.55 eV) semiconductor with P-S σ bonds. The electronic properties of P$_2$S$_3$ structure can be modulated by stacking into multilayer P$_2$S$_3$ structures, forming P$_2$S$_3$ nanoribbons or rolling into P$_2$S$_3$ nanotubes, expanding its potential applications for the emerging field of 2D electronics.

**Text:**

The epic discovery of graphene [1], a two-dimensional (2D) phase of carbon, has paved the way for the synthesis of many other 2D materials, including the 2D insulator boron nitride (BN) [2–4], graphene-like group IV 2D materials, i.e. semimetallic silicene, germanene, and stanine [5–11], 2D transition-metal dichalcogenides [12–16], such as molybdenum disulfide [2,17,18] and tungsten disulfide [19], and recently, 2D phosphorus, i.e. phosphorene [20], which holds great promise for applications in electronics and optoelectronics.

The reduced dimensionality and symmetry of 2D materials, lead to unique electronic, optical and mechanical properties that differ from their bulk counterparts [21,22], offering possibilities for numerous advanced applications. For instance, transistors made with single layer $MoS_2$ present room-temperature current on/off ratios of $10^8$ [18]. Two-dimensional materials also offer novel opportunities for fundamental studies of unique physical and chemical phenomena in 2D systems [23,24]. More interestingly, stacking different 2D crystals into hetero-structures (often referred to as 'van der Waals') have recently been achieved and investigated, revealing new phenomena and novel properties [25].

Over the past decade, a number of experimental methods have been developed to produce monolayer nanosheets by exfoliating layered materials with oxidation, ion intercalation/exchange, or surface passivation induced by solvents [26,27]. Theoretical methods have been employed to search new two-dimensional materials such as evolutionary crystal structure search [28–30] and particle swarm optimization (PSO) [31]. For example, a novel 2D $Be_2C$ monolayer, with each carbon atom binding to six Be atoms to form a quasi-planar hexacoordinate carbon moiety, was discovered by Li *et al* [32]. And in our previous work [33], we discovered a novel light-emitting 2D crystal with a wide direct band gap, namely $S_3N_2$ monolayer, using the evolutionary crystal structure search method. The amazing properties of the $S_3N_2$ 2D crystal inspired us to explore the possibility of other group V-VI 2D crystals.

In this work, we proposed a new two-dimensional crystal, diphosphorus trisulfide ($P_2S_3$) with 3 polymorphs: $α$-$P_2S_3$ (Fig. 1(a)), $β$-$P_2S_3$ (Fig. 1(b)) and $γ$-$P_2S_3$ (Fig. 1(c))). The ground state structure of these polymorphs were obtained using the evolutionary algorithm driven structural search code USPEX [28–30]. These $P_2S_3$ polymorphs were further geometry optimized with density functional calculations with Perdew–Burke–Ernzerhof (PBE) [34] exchange-correlation functional using the Cambridge series of total-energy package (CASTEP) [35,36]. A plane-wave cutoff energy of 700 eV was used, and Monkhorst-Pack [37] meshes with 0.02 Å$^{-1}$ *k*-point spacing were adopted. The convergence test of cutoff energy and *k*-point mesh was conducted. In calculating the binding energy of bilayer $α$-$P_2S_3$, the empirical dispersion correction schemes proposed by Grimme (D2) [38] was used in combination with PBE functional to properly describe the van der Waals (vdW) interactions between $α$-$P_2S_3$ layers. Since the band gaps may be dramatically underestimated by the GGA level DFT [39,40], the quasiparticle GW calculation [41] of the band structure was carried out using YAMBO software package [42]. The Green function and Coulomb screening were constructed from the PBE results obtained by Quantum Espresso [43], and the plasmon-pole model was used for computing the screening. The $G_0W_0$ approximation was adopted in carrying out the GW approximation, since it gives very good results for many materials without $d$ electrons [44]. All structure optimizations were conducted without imposing any symmetry constraints. The conjugate gradient method (CG) was used to optimize the atomic positions until the change in total energy was less than $5 \times 10^{-6}$ eV/atom, maximum stress within 0.01 GPa and the maximum displacement of atoms was less than $5 \times 10^{-5}$ Å.

The fully relaxed $P_2S_3$ polymorphs are depicted in Fig. 1. These $P_2S_3$ polymorphs are 2D covalent networks composed solely of σ bonds (bonding is depicted by isosurfaces of the electron density). For α-$P_2S_3$ (space group Pmn21), the unit cell (see Fig. 1(a)) consists of ten atoms with lattice constants $a$ = 4.71 Å, $b$ = 10.62 Å, P-S bonds with bond lengths $d_1$ = 2.14 Å, $d_2$ = 2.12 Å, $d_3$ = 2.15 Å, and bond angles $\theta_1$ = 103.8°, $\theta_2$ = 105.7°, $\theta_3$ = 96.2°, $\theta_4$ = 94.4° and $\theta_5$ = 107.4°. The unit cell of β-$P_2S_3$ (space group Cmm2) consists of five atoms with lattice constants $a_1$ = $a_2$ = 5.35 Å, the angle between unit vector $a_1$ and $a_2$, $\gamma$=108.8°, P-S bonds with bond lengths $d_1$ = 2.14 Å, $d_2$ = 2.15 Å, and bond angles $\theta_1$ = 93.0°, $\theta_2$ = 111.4°, $\theta_3$ = 132.4° and $\theta_4$ = 94.1° (see Fig. 1(b)). The unit cell of γ-$P_2S_3$ (space group P31m) consists of five atoms with lattice constants $a_1$ = $a_2$ = 5.92 Å, P-S bonds with bond length $d_1$ = 2.16 Å, and bond angles $\theta_1$ = 95.1° and $\theta_2$ = 104.9° (see Fig. 1(c)). For these polymorphs, the Brillouin zones with the relevant high-symmetry k-points are depicted in the inset figures in Fig. 1. The cohesive energies of α-$P_2S_3$, β-$P_2S_3$ and γ-$P_2S_3$ are -3.64 eV, -3.59 eV and -3.60 eV, so the most energetically stable polymorph is α-$P_2S_3$.

Even if the structure optimizations using CG method indicate the stability of these free standing $P_2S_3$ polymorphs, further tests should be conducted to assure these polymorphs are stable in the local minimum and can remain stable above the room temperature. First, by conducting phonon dispersion calculation of the free-standing P$_2$S$_3$ polymorphs, we verified that all phonon frequencies of α-P$_2$S$_3$ are real (Fig. 2(a)), confirming the dynamic stability of this structure. However, β-P$_2$S$_3$ and γ-P$_2$S$_3$ are not dynamically stable, since they have imaginary phonon frequencies (Fig. 2(b-c)). Thus, our focus will be on the properties of the dynamically stable α-P$_2$S$_3$ in the following discussions.

For α-P$_2$S$_3$, the enthalpy of formation Δ$H$ from the elements

$$\alpha\text{-P}_2\text{S}_3 = 2\,\text{P (s)} + 3\,\text{S (s)} \quad (1)$$

calculated by CASTEP at T=0 K is -14.2 kcal/mol. This means α-P$_2$S$_3$ is an energetically favorable composition relative to phosphorus and sulfur in their solid state.

To verify the stability of the structure under high temperatures, *ab initio* molecular dynamics (MD) simulations (shown in Fig. 3) were performed at the PBE [34] /GTH-DZVP [45] level in the NPT ensemble with the CP2K [46] code. The simulations were run for 10 ps under ambient pressure under temperature T= 1000 K and no breaking of the bonds was seen, indicating a high degree of stability. To further verify the chemical stability of the structure in air, *ab initio* MD of α-P$_2$S$_3$ crystal exposed to very high pressure gases (O$_2$, N$_2$, H$_2$O and H$_2$) at temperatures T= 1000 K were conducted (Fig. 4). In our MD simulations, the number density of gas molecules was 57.5 × 10$^{25}$ m$^{-3}$. Such high gas pressure were also used to study oxidation of graphene [47] and phosphorene [48] with MD simulations. α-P$_2$S$_3$ structure remains intact under these very high gas pressure for 10 ps (Fig. 4), indicating its chemical stability in air above the room temperature.

The quasiparticle and DFT band structures and density of states of the 2D α-P$_2$S$_3$ crystal are shown in Fig. 5. Calculations carried out using GW method showed that the α-P$_2$S$_3$ structure is a semiconductor with a wide band gap of 4.55 eV (calculations carried out using PBE functional underestimate the band gap by 2.05 eV). The valence band maximum (VBM) is composed of mainly the orbitals of sulfur atoms, while the conduction band minimum (CBM) is more or less evenly contributed by the orbitals of phosphorus and sulfur atoms (see Fig. 5).

Our analysis shows that not only single-layer $\alpha$-$P_2S_3$, but also bilayer and its 3D phase constructed by the stacking of $\alpha$-$P_2S_3$ monolayers, are stable. The minimum energy stacking for the bilayer and 3D phase are shown in inset figures in Fig. 5. The binding energy between layers is weak and is only 0.13 J/m$^2$, which is predominantly vdW attraction energy. The DFT band gaps are reduced by 0.14 eV by just stacking $P_2S_3$ into a bilayer. By stacking $P_2S_3$ into 3D $P_2S_3$ crystal, the DFT band gap is further reduced to 2.18 eV. Beside stacking, the electronic properties of $P_2S_3$ can be modulated by cutting into $P_2S_3$ nanoribbons or rolling up to form $P_2S_3$ nanotubes, expanding its potential applications in 2D electronics.

In conclusion, we predicted a novel two-dimensional trisulfur dinitride ($P_2S_3$) crystal with high thermal stability using *ab initio* simulations. Band structures calculated using the GW method indicate that 2D $P_2S_3$ crystal is a semiconductor with wide band gap of 4.55 eV. The $P_2S_3$ solid is the first 2D crystal composed of phosphorus and sulfur, which also forms stable bilayer, 3D layered solid and nanoribbon structures. These structures with tunable band structures shed light on the applications for the emerging field of 2D electronics.

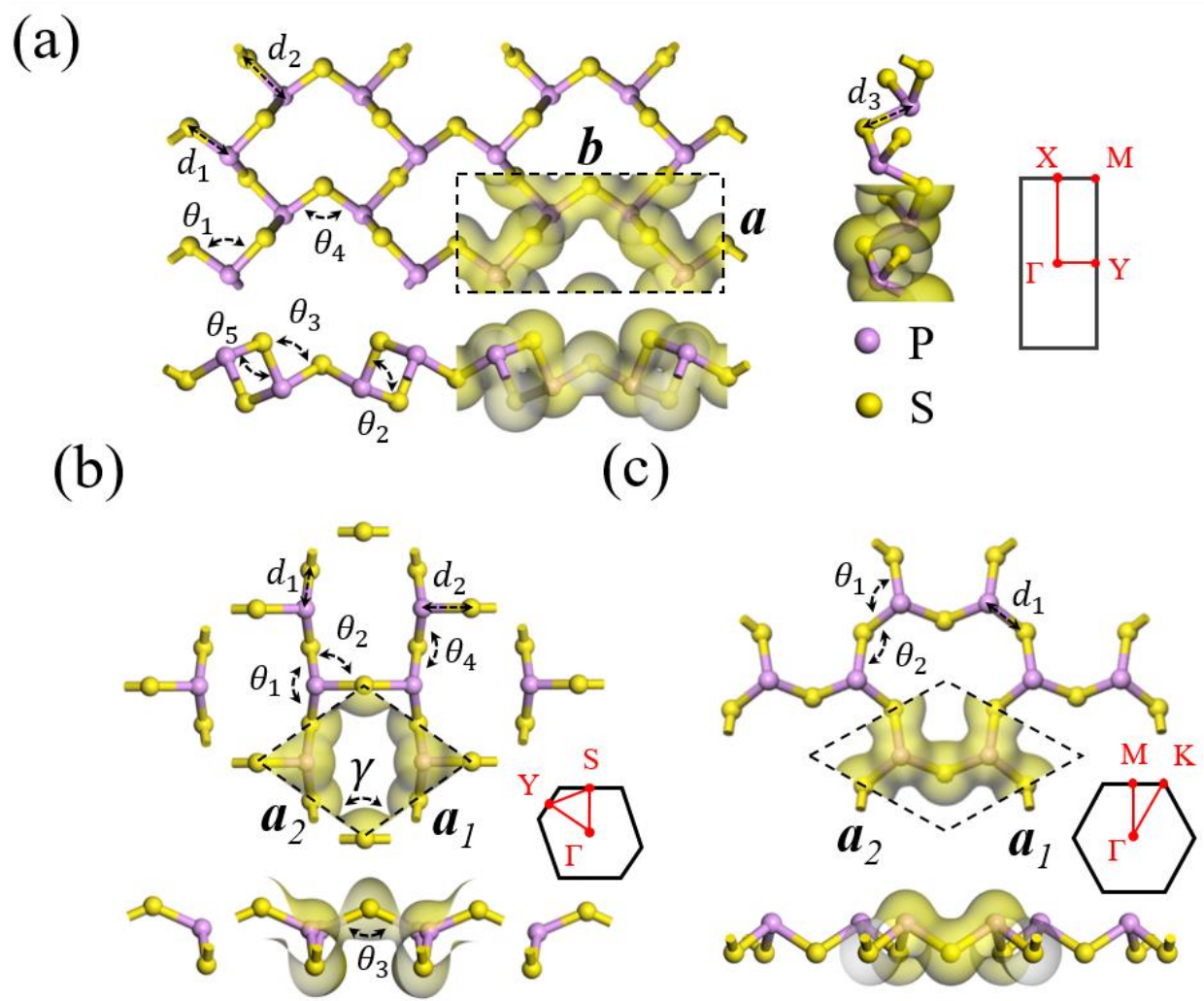

**Fig. 1.** 2D crystalline structures of *α*-P$_2$S$_3$ (a), *β*-P$_2$S$_3$ (b) and *γ*-P$_2$S$_3$ (c). The Brillouin zone of each polymorph, with the relevant high-symmetry *k*-points indicated, is depicted in the inset figure. Bonding is depicted by an isosurface of the electron density.

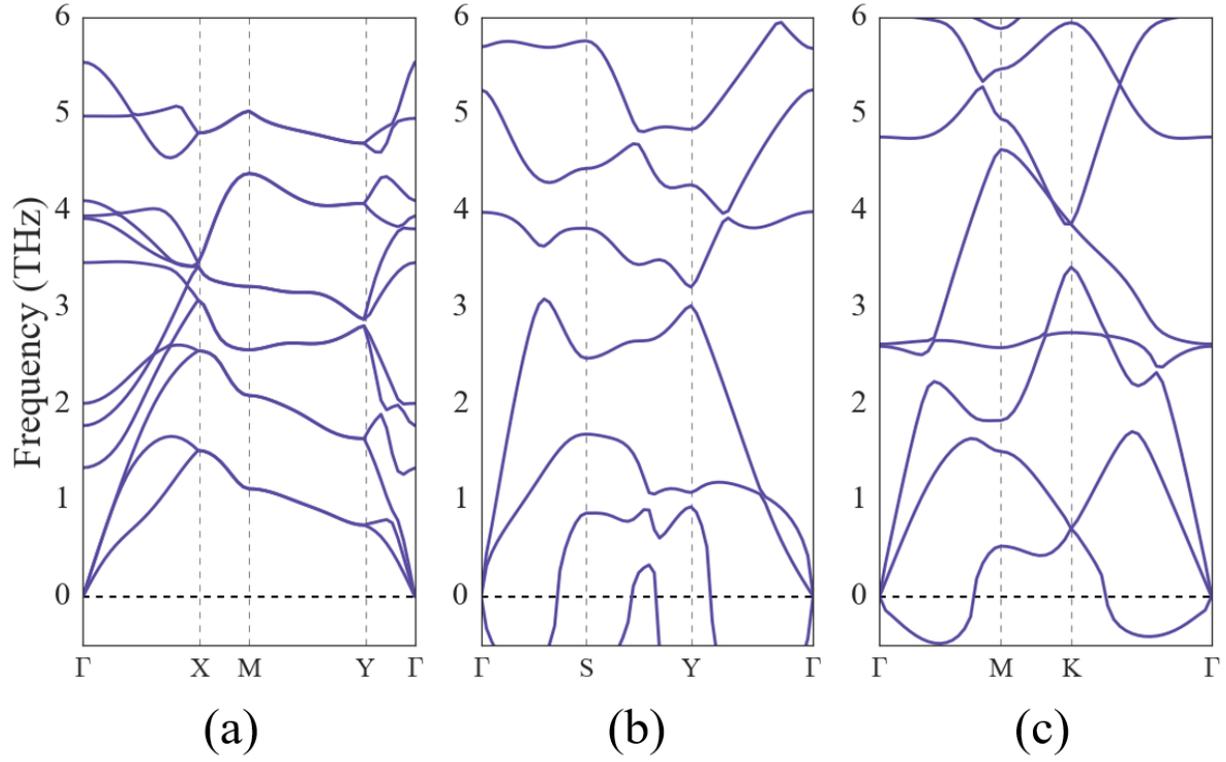

**Fig. 2.** The phonon dispersion relations of α-$P_2S_3$ (a), β-$P_2S_3$ (b) and γ-$P_2S_3$ (c). Dynamic stability of α-$P_2S_3$ is indicated by the absence of negative frequencies. However, β-$P_2S_3$ and γ-$P_2S_3$ are dynamically unstable, due to the presence of imaginary phonon frequencies.

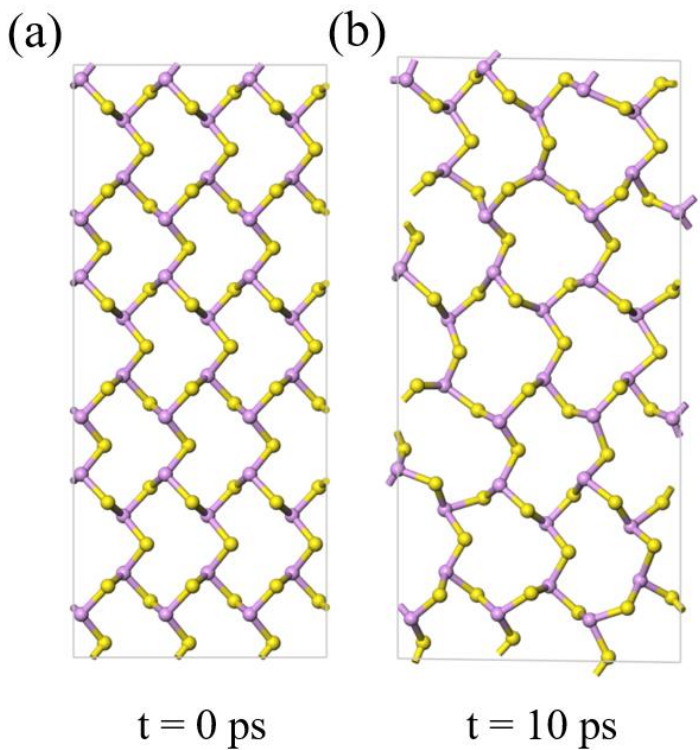

**Fig. 3.** *Ab initio* MD snapshots of the α-P$_2$S$_3$ supercell structures at temperature T = 1000 K under ambient pressure at time t = 0 ps (a) and t = 10 ps (b). No breaking of the bonds was seen during the 10 ps *ab initio* MD simulation, indicating a high degree of thermal stability.

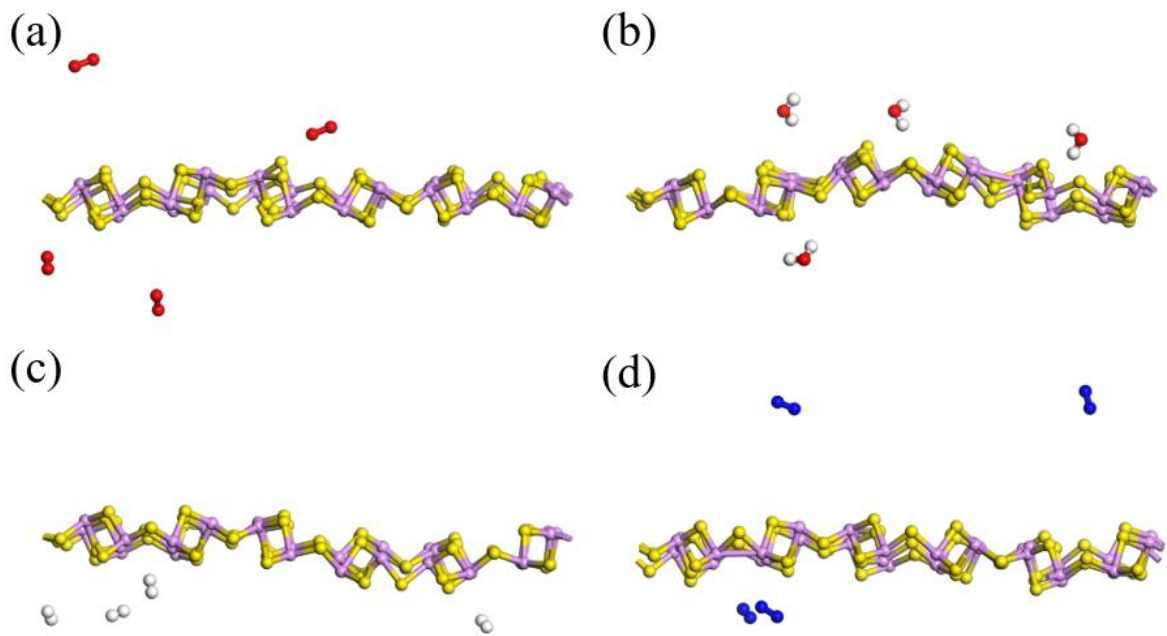

**Fig. 3.** *Ab initio* MD snapshots of the α-$P_2S_3$ supercell structures exposed to the high pressure oxygen gas (a), water vapour (b), hydrogen gas (c), and nitrogen gas (d) at temperatures T = 1000 K. $P_2S_3$ structures remain chemically stable during these 10 ps simulations, confirming the chemical stability of α-$P_2S_3$.

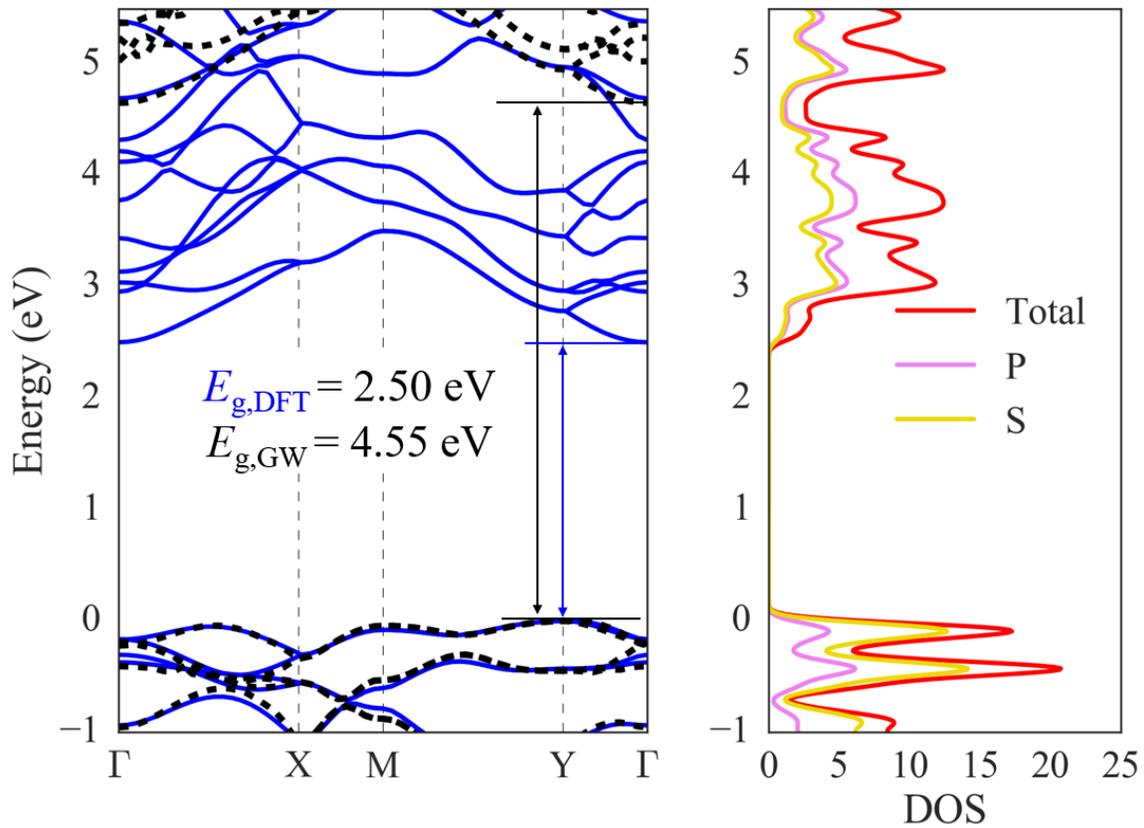

**Fig. 4.** Calculated band structure (left) obtained with the PBE functional (blue lines) and the GW method (black dash lines) for the α-$P_2S_3$ solid. The DOS (right) is obtained with the PBE functional. α-$P_2S_3$ crystal is a semiconductor with a wide band gap of 4.55 eV. The valence band maximum (VBM) is composed of mainly the orbitals of sulfur atoms, while the conduction band minimum (CBM) is more or less evenly contributed by the orbitals of phosphorus and sulfur atoms.

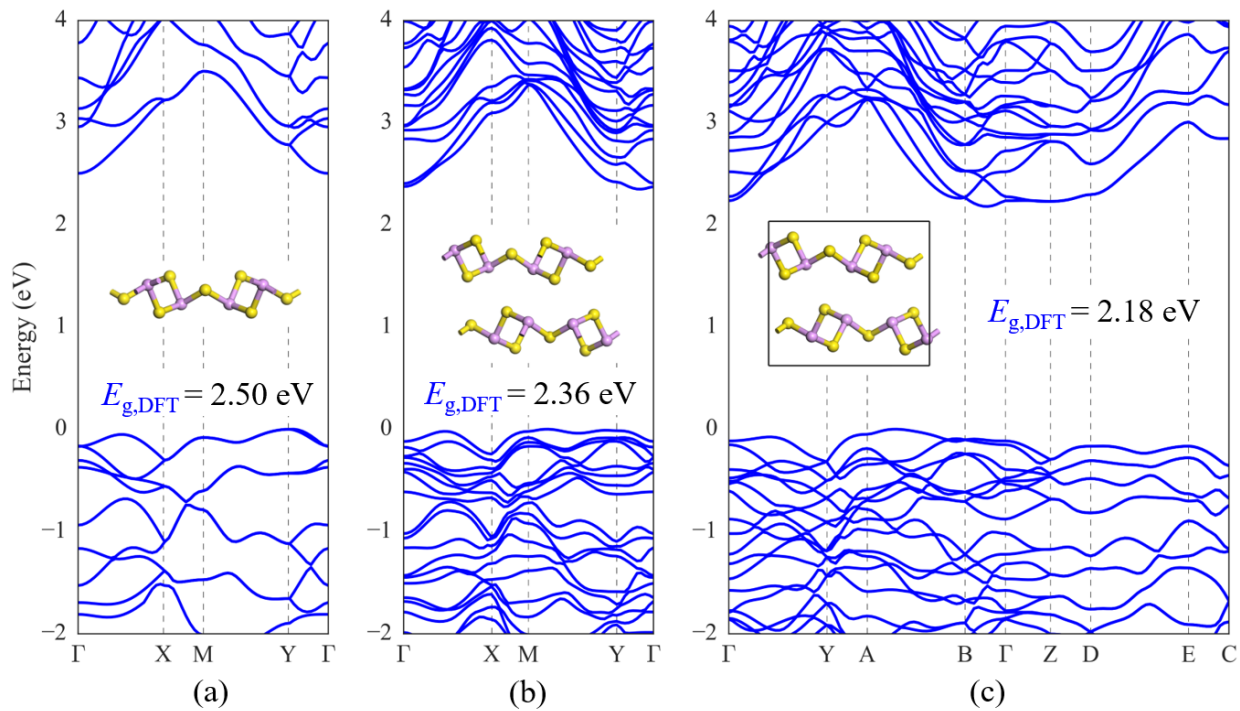

**Fig. 5.** The electronic band structures of the α-P$_2$S$_3$ monolayer (a), α-P$_2$S$_3$ bilayer (b) and α-P$_2$S$_3$ 3D crystal, obtained with the PBE functional. Monolayer, bilayer and 3D crystal structures of α-P$_2$S$_3$ are shown in inset figures.

AUTHOR INFORMATION

**Corresponding Author**

Xi Chen, E-mail: xichen@columbia.edu

**Author Contributions**

These authors contributed equally.

**Notes**

The authors declare no competing financial interests.

ACKNOWLEDGMENT

The authors acknowledge the support from the National Natural Science Foundation of China (11372241 and 11572238), ARPA-E (DE-AR0000396) and AFOSR (FA9550-12-1-0159).